# Introducing Business Language Driven Development

Rogerio Atem de Carvalho, Rodrigo Soares Manhaes, Fernando Luiz de Carvalho e Silva

Emails: ratem@iff.edu.br, rmanhaes@iff.edu.br, fernando.carvalho@iff.edu.br
Nucleo de Pesquisa em Sistemas de Informação (NSI), Instituto Federal Fluminense (IFF),
R. Dr. Siqueira, 273, Campos/RJ, Brazil

**Abstract.** A classical problem in Software Engineering is how to certify that every system requirement is correctly implemented by source code. This problem, albeit well studied, can still be considered an open one, given the problems faced by software development organizations. Trying to solve this problem, Behavior-Driven Development (BDD) is a specification technique that automatically certifies that all functional requirements are treated properly by source code, through the connection of the textual description of these requirements to automated tests. However, in some areas, such as Enterprise Information Systems, requirements are identified by Business Process Modeling – which uses graphical notations of the underlying business processes. Therefore, the aim of this paper is to present Business Language Driven Development (BLDD), a method that aims to extend BDD, by connecting business process models directly to source code, while keeping the expressiveness of text descriptions when they are better fitted than graphical artifacts.

## l. INTRODUCTION

One of the classical Software Engineering problems is how to guarantee that all functional requirements are implemented correctly by the source code. Many techniques were developed such as the construction of the so-called Requirements Traceability Matrix [12], which keeps track of the source code elements that actually realize the requirements. In that way, developers should determine "manually", for each functional requirement, which classes and methods implement it. This technique is time consuming and error prone, basically because the traceability work is done in an artificial way, through the introduction of an additional artifact: the matrix. The ideal way of keeping track of requirements is by directly joining them to source code.

**Enters BDD**

Following this reasoning, Behavior-Driven Development (BDD) [1] is a specification technique that automatically certifies that all functional requirements are treated properly by source code, through the connection of the textual description of these requirements to automated tests. BDD relays heavily on Test-Driven Development (TDD) [2], which in turn is a technique that consists of writing test cases for any programming task (new or adapted feature, improvements, bug corrections etc), before these implementations are performed. According to Koskela [3], TDD is intended for "solving the right problem right", meaning to achieve the correct solution that exactly matches the business problem.

BDD starts with textual descriptions of the requirements using specific keywords that tag the type of sentence, indicating how the sentence is going to be treated in the subsequent development phases. These descriptions are written in accordance to a given language that is able to represent requirements in a way easily comprehensible by both users and developers, the so-called Ubiquitous Language (UL), which is a language structured around the domain model and used by all team members to connect all the activities of the team with the software [10] .

**Enters BLDD**

Although using text may be the more natural way of describing a system's requirements, and besides the fact that BDD provides the structure for doing so, in many cases, in special during the development of Enterprise Information Systems (EIS), Business Process Modeling (BPM) takes place, providing models in different notations, such as UML Statechart Diagrams or different Petri Nets flavors.

Therefore, the direct connection of business process models to the code that implements them seems to be a solution for the problem of obtaining low-cost, high-confiability requirements tracing. Although one can think the natural way of doing that is through the use of workflow engines, those engines usually are dependent of a given notation and underlying platform, while the method here proposed is independent of any of these supportive tools, although can work with them[1].

This paper aims to introduce Business Language Driven Development, or BLDD for short, as a way of developing high quality business systems. BLDD intends to use BDD's philosophy and tools, and complement it by adding the Enterprise Information Systems' Ubiquitous Language: Business Process Models. Text descriptions of requirements are still used, for those requirements for which doesn't make sense to use process models, such as reports. Also, acceptance conditions are better described using texts.

This work is a natural development of the original ideas presented by [4] and [5], and is organized as follows: after this introduction, BDD main features are briefly presented; followed by the focus of this paper, which is to introduce BLDD; finally some conclusions and future work are presented.

---

[1] In fact, the authors are developing a structure to use BLDD into the ERP5 system. This system is implemented on top of the Zope framework, which uses the DCWorkflow state-based workflow engine.

## 2. A Brief Introduction to BDD

This topic aims at introducing BDD, however, since the focus of this paper is on present the areas where BPDD complements BDD, many details of the BDD process will not be treated here.

BDD starts with the identification of a business requirement using a set of specific tags, forming a simple Ubiquitous Language (UL). The requirement is described according to the following template [6]:

> As a **Role**
> I request a **Feature**
> To gain a **Benefit**

This template is used basically to make the business value of the requirement explicit for developers and users. Following this identification, a series of possible scenarios for the requirement must be written, by describing the scenarios as sets of Given-When-Then constructs:

> Given a **Context** (or a system **State**)
> When an **Event** happens (or an user **Action**)
> Then an **Action** is taken (or a system **Reaction**)

From this point onwards, a tool is used to parse the scenarios and map the natural language sentences into the underlying programming language equivalent calls, while keeping the same abstraction level, as the following feature exemplifies:

**Feature:** Manage Budget
  **In order to** Create a Budget
  **As a** Vendor
  **I want to** Add products to Budget

  **Scenario**: User adds a new Bid
    **Given** I go to the new Bid page
    **And** I fill in "Client" with "My Client Name"
    **And** I fill in "Product" with "XXXXXX"
    **And** I fill in "Quantity" with "1"
    **When** I press "Add"
    **Then** I should be on the Budget list page
    **And** I should see "Test Product XXXXXX"

After the scenario "User adds a new Budget " is parsed, the calls in Code 1 will be generated:

    @step(r'Given I go to the new Budget page ')

```
def given_i_go_to_the_new_budget_page (step):
    # code goes here
@step(r'I fill in "Client" with "My Client Name" ')
def given_i_fill_in_Client_with_my_client_name (step):
    # code goes here
@step(r'I fill in "Product" with "XXXXXX" ')
def given_i_fill_in_product_with_XXXXXX (step):
    # code goes here
@step(r'I fill in "Quantity" with "1" ')
def given_i_fill_in_quantity_with_1 (step):
    # code goes here

@step(r'When I press "Add" ')
def when_i_press_add(step):
    # code goes here

@step(r'Then I should be on the Budget list page ')
def then_i_should_be_on_the_budget_list_page (step):
    # code goes here
@step(r'I should see "Test Product XXXXXX" )
def then_i_should_see_test_product_XXXXXX(step):
    # code goes here
```

Code 1: Generated steps

With the generated code, the programmer can then write tests that will drive all the design and implementation of the system, by using TDD, which prescribes automated unit tests, pieces of code that excite the code that has to be implemented. By wrapping all implementation code with tests, which in turn are automatically tied to the business requirements using the Given-When-Then (GWT) constructs, fulfills the needs for the documentation of requirements in most projects, having the advantage of making the whole system verifiable at any time.

Using BDD allows reducing the risks and effort to implement a given change in an information system; therefore the system can be continuously improved without to fall into the famous Boehm's cost of change curve, which established that the cost of change in a software project increases exponentially through the time [8]. Moreover, the THEN constructs represent the acceptance criteria, giving an objective method to state that a given requirement is "done".

BDD provides an automated and cost effective way of keeping requirements traceability, addressing each of the challenges introduced by [13]:

– Cost: it is able to provide full tracing in a way that is even cheaper than Value-based requirement tracing [14]. Requirements are tied to tests, if tests return non-expected values or simply are not present, the tool will automatically point out the problem.

- Managing change: there is no need to impose "strong discipline in maintaining the accuracy of traceability", instead, BDD tools make the checks automatically. If a requirement changes, the tests will not run until the code is also changed accordingly. Moreover, by changing a requirement and immediately running a complete build on top of the tests, errors will pop-up in every place where the system must be changed, easing effort estimation.
- Different stakeholder viewpoints: according to [13] standards provide no guidance to traceability. Through BDD, it is possible for any stakeholder to check system consistency, since it is based on executable documentation. Even end users can push a button and check the error messages.
- Organizational problems: according [13], "the easiest way to correct organizational problems related to traceability is through the use of policy and training". BDD provides a proper policy for traceability because it enforces the connection of code to the requirements. Thus, it is a question of training for BDD.
- Poor tool support: BDD offers a tool for each abstraction level, in fact, there is no way of doing wrong things. An error will occur if a test is missing, or if it results in different values, all phases are automated and well connected.

For a BDD developer the considerations above can even appear nonsense, since the correct use of its tools an methods will avoid the occurrence of all these problems.

**Mapping from Business Process Representations to BDD**

According to [7], the Given-When-Then convention connects the human concept of cause and effect, to the software concept of input/process/output. Also according to [7] this convention "is simply a state transition, and that BDD is really just a way to describe a finite state machine. Clearly "GIVEN" is the current state of the system to be explored. "WHEN" describes an event or stimulus to that system. "THEN" describes the resulting state of the system." Therefore, the set of all scenarios of a given business requirement can be represented by a Finite State Machine, and, as a consequence, as a Petri Net. Although there are lots of business process (or workflow) patterns, as described by [9], they are all formed by basic constructs. Thus, by mapping these basic constructs, in any notation, to Given-When-Then (GWT) constructs, its is possible to build any of these patterns using BDD Ubiquitous Language. This mapping is provided by [5], and basically is used to show that from the main business process notations it is possible to reach GWT constructs, consequently, it is possible to substitute the BDD's textual constructs by business process models.

## 3. Business Language Driven Development

The central idea proposed by BLDD is to complement GWT constructs with some Business Process Modeling notation, such as Petri Nets or UML Activity Diagrams. In other words, to use a BPM notation as (part of) the Ubiquitous Language. In this way, instead of running textual specifications for checking for system consistency, a representation of the business process is run. Moreover, this run can be performed step-by-step, with the final user choosing which path to follow among the various present in a given process. For each user's interaction with the representation, a given set of test is activated, exciting the corresponding system code, and making the appropriate user interface to pop, with the test values already filled in the proper places. The user then can verify if the system is behaving in accordance with its requirements by watching a live demonstration of the business processes implemented by the information system.

**A Composition of Languages?**

We believe that the Ubiquitous Language can be a composition of text, graphical artifacts, as well as mathematical formalisms, given that, depending on the complexity of the system and its knowledge area, a single language cannot represent its requirements completely. For instance, a given business process may use mathematical tools to support decision making, therefore, it is interesting to use the exact language on both the system representation and its implementation through code. The idea is keeping the code the closest possible to the description.

While BDD omits details of its language – for instance, how to represent parallelism, for BLDD the UL is a composition of graphical artifacts representing the business process and textual information. Also, BDD considers that GWT is "a ubiquitous language for system analysis", we prefer to think in terms of "a composition of languages to represent the domain area." In fact, this second definition applies more to what we call Shared Language Driven Development. Therefore, it is understood in this proposal that:
a) when referring to the general idea of using a composition of languages, previously agreed by stakeholders, it is going to be called Shared Language Driven Development (SLDD);
b) when referring for the specific case of using BPM plus a domain specific language (DSL) as an input for the BDD machinery, it is going to be called Business Language Driven Development (BLDD), the focus of this article.
In other words, SLDD is a generalization of BDD, and BLDD is an extension to BDD that uses BPM.

**Introducing the BLDD Proof of Concept[2]**

---

[2] This tool was first shown in the blog http://eis-development.blogspot.com

Executable Documentation is a vital element for achieving end to end testing and thus real code quality (instead of "quality by annotation"). The core idea is to use some Business Process (BP) modeling tool to run, step by step, a given process representation and, at each step, excite the associated automated tests, making the system to pop-up accordingly as the user clicks on the representation. It is the same thing we see when we use tools such as Selenium, however, in that case, we would run the tests step by step and, instead of using textual descriptions of the requirements (like in BDD), we use a BP Model. It is known that there isn't a real standard for Business Process Modeling, in fact there are various notations, such as UML Models, Petri Nets, BPMN, derivations of these etc. It is also known that many EIS are based on Workflow Engines, and each of these engines has its particularities. Therefore, this proposal can be implemented in two different ways:

a) Use a generic BP representation: the choice was Petri Nets (PN), because one can derive others representations from them, such as Statechart and Activity Diagrams, and also there is the "Token Game", which is the simulation of the BP execution.

b) Use some workflow engine to work on top of it. The first choice was working with ERP5 [15] an open source ERP (FOS-ERP), based on Zope, which in turn uses a state-based workflow engine, DCWorkflow.

This paper will focus now on (a), given that (b) is still in development. The first step for implementing a proof of concept was to use some Petri Nets open source tool that would be able to run Token Games, so that models can be animated – being Woped [16] our choice.

Using a Woped's example PN, a very basic application was created, without taking much care of the application logic or GUI - the goal is to show the proposal, not discuss BP modeling or GUI implementation. The screenshot in Figure 1 shows the moment when the user reaches an OR-Split and must choose which path to follow in the workflow.

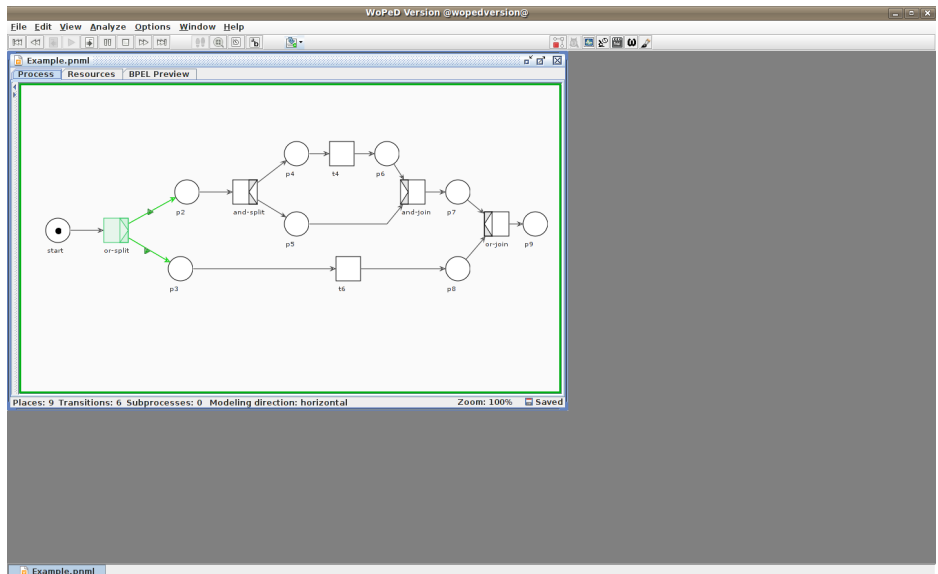

Figure 1: An user reaches an OR-Split in the example process

The OR-Split in Figure 1 is implemented as a list box. Appropriate tests excite this code, as shown in Figure 2.

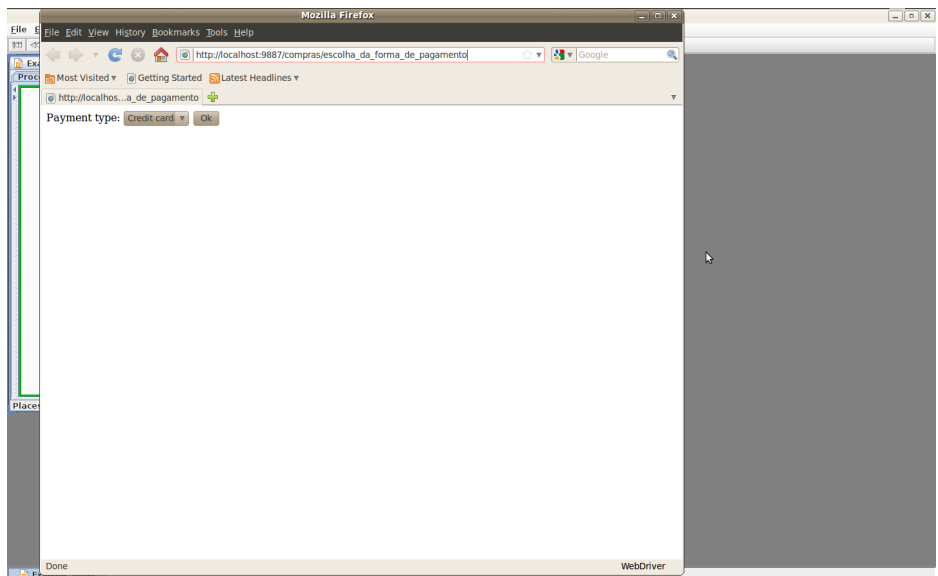

Figure 2: The equivalent test run for the OR-Split

Given that tests "are by example", the user does not interact with the real system, instead, he/she is a spectator of the tests, launching them by choosing which path to follow in the PN. After checking if the system is implemented in the expected way, the user closes the GUI and then choose the path equivalent to Credit Card, making the test code to run this option, which, in turn, makes the form in Figure 3 to pop. Meanwhile in Woped, the user can see a message that shows which scenario was chosen, including the Given-When-Then equivalent instructions (Figure 4).

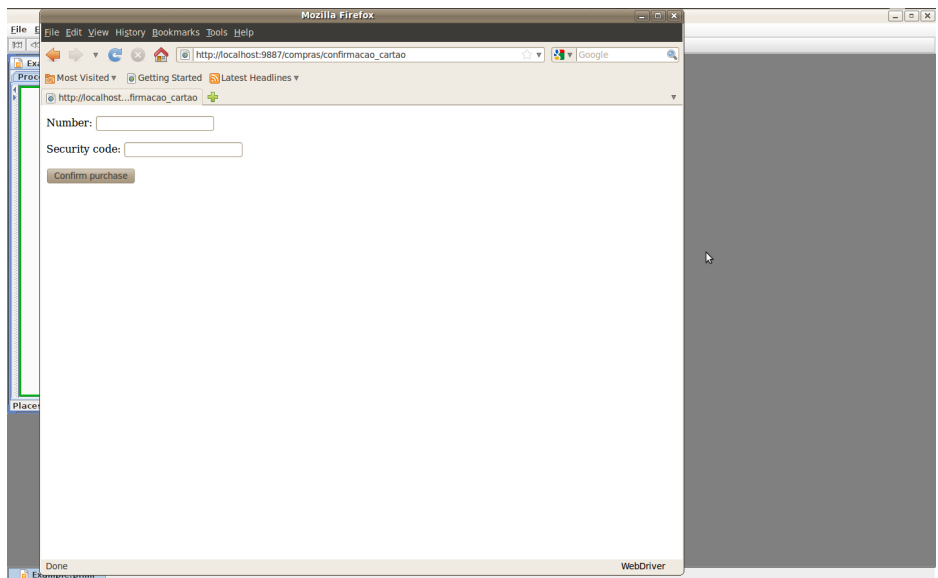

Figure 3: Credit Card data form

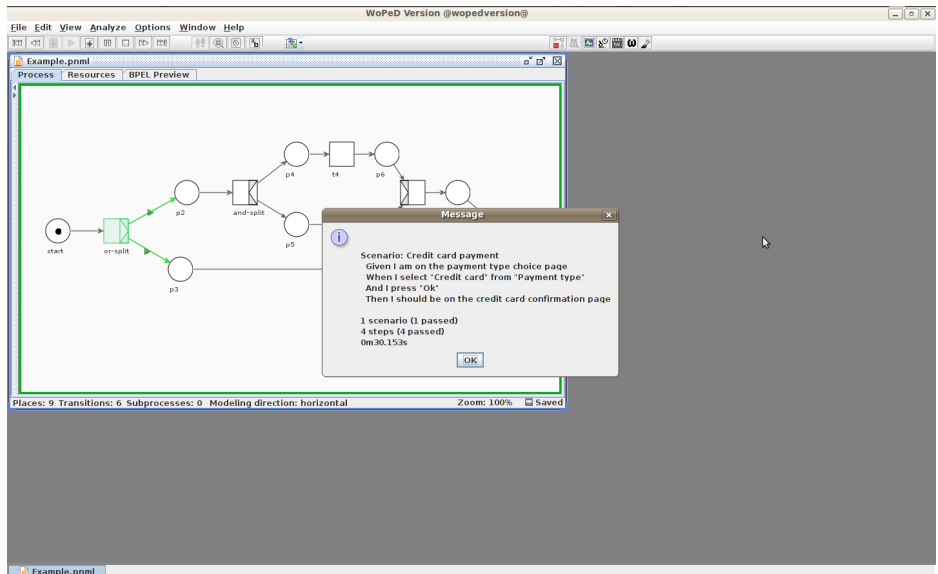
Figure 4: Message with equivalent GWT for the upper flow.

When the user clicks on OK, the flow goes to the next step, an AND-Split (Figure 5) and the tests responsible for filling the form will run, as shown in Figure 6.

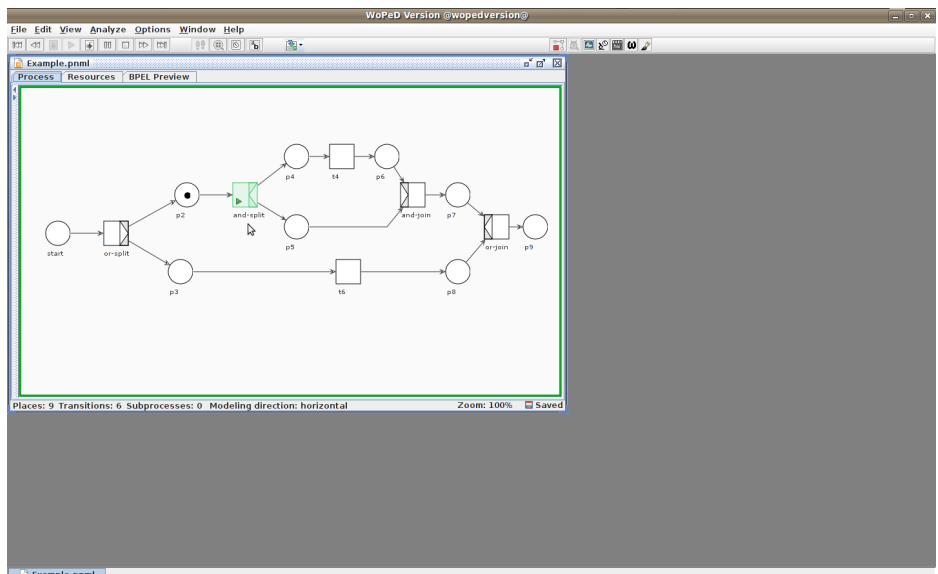
Figure 5: The process enters an AND-Split

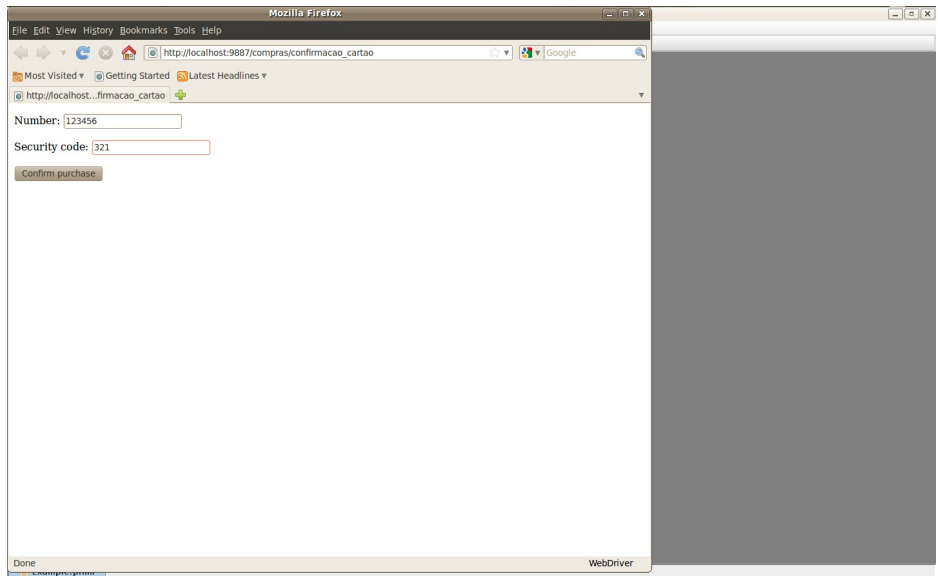

Figure 6: Running the test for filling Credit Card data

One problem with BDD's UL is that constructions like parallel threads are not clearly represented, they are simply represented by ANDs in the Then clauses, which can poorly represent concurrent tasks. This type of problem is clearly shown by the mapping from UML Statechart Diagrams' constructs to BDD's GWT [5].

Continuing with the example, the process now reached an AND-Split, making the tests automatically fill the form with the Credit Card data, and driving the validation process to stop at the AND-Split, waiting for the user to click on it. When the user clicks on it, Figure 7 shows the equivalent GWT construct, being the "Wait for confirmation..." part equivalent to the upper flow of the split.

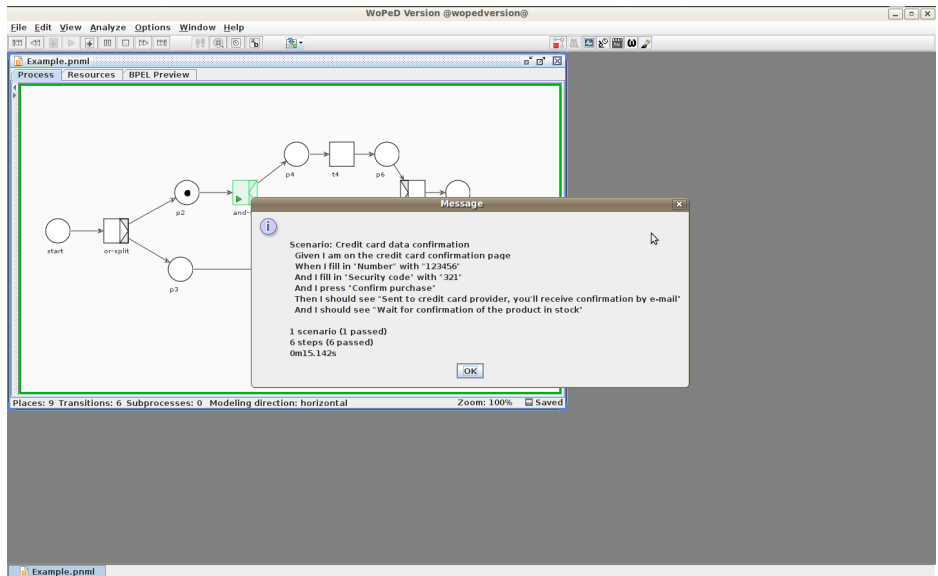

Figure 7: The equivalent GWT construct for the AND-Split

As the user clicks OK, two subsets of tests excite in parallel two different parts of the system: the one that tests the sending of an email (the lower flow of the split, not shown here since there is no user interaction in it) and the inventory-checking procedure, of which equivalent in GWT is presented by Figure 8. After the user clicks OK, the process will go forward, reaching an OR-Join, as shown in Figure 9.

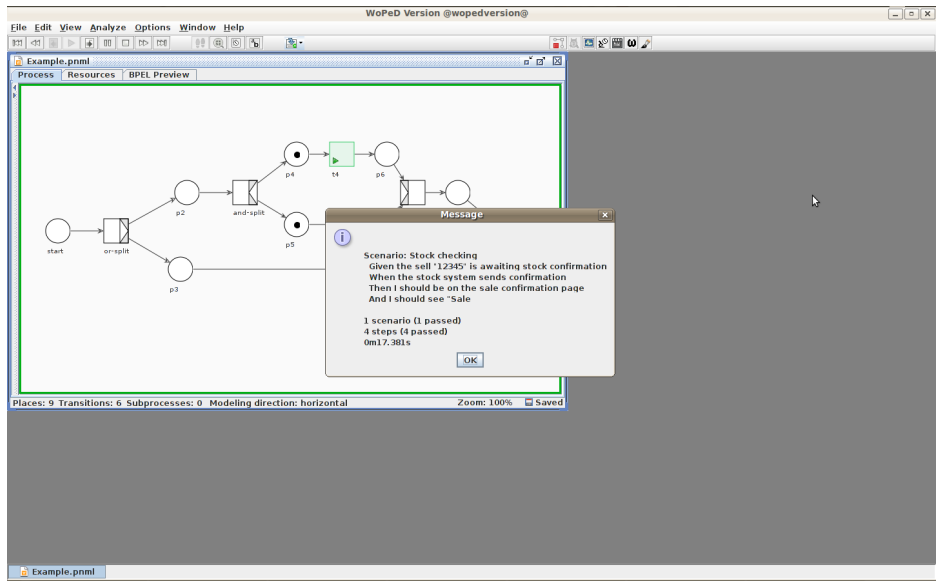

Figure 8: Excitation of inventory-checking procedure

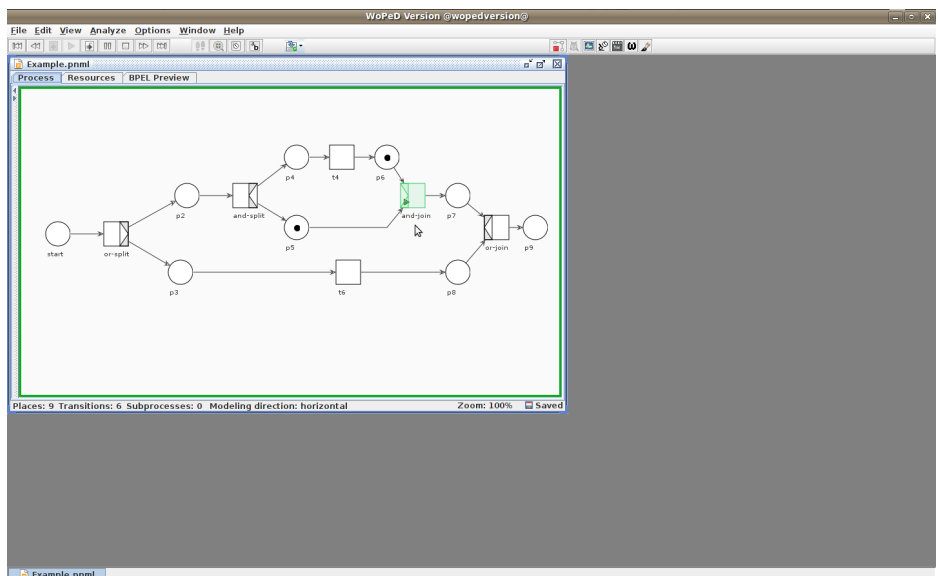

Figure 9: An OR-Join is reached

Since BDD tools are used for implementing this proof of concept, GWT constructs were written "by hand", however, each BP representation must have a specific language for accompany it. For instance, for Statechart Diagrams, tags State-Event-Action may be used to map from the graphical representation to the test set, as shown in Code 2.

```
@SourceState('current state')
# excite the code that treats the exit of the current state
@Event ('event')
# excite the code that treats the event
@Transition('produces some outcome')
# excite the code that produces the expected outcome
@DestinationState('new state')
#excite the code that treats the entrance into a new state
```
<div align="center">Code 2: Possible tags for state-based BP</div>

It is important to note that specific logic for each of the PN structures, such as AND and OR splits, was implemented, having the Keep It as Simple as Possible (KISP) principle in mind. Also, it is not proposed to substitute GWT constructs, but to complement them, because simpler requirements can be described without the need of BP models.

An interesting possibility is to generate the test calls from the business process model, initially returning a dummy message that indicates that the test and the acceptance criterion need to be implemented and defined respectively. By accessing the BP representation, the tests are called, and if not implemented correctly the system will not pop correctly.

**Error Handling**

One way of making a tool prove its utility is using it in scenarios where things go wrong. Figure 10 shows what happens if the code implemented is not in compliance with the requirements. The user will see a message saying that the scenario didn't pass.

In the message presented in Figure 10, the error was generated because the last step was not implemented correctly: the message "9 sales awaiting to be sent" was not generated by the system, therefore, the developer must check the code and correct it. In that way it is possible to immediately identify where the system must be improved to work as expected, which is very useful when it is still in development, or when changes are needed.

In that last case (change), one can imagine that the "9 sales awaiting to be sent" message represents an improvement to this business process. If it is inserted into the specifications and the simulation is ran, the exact point where the system must be modified to become compliant to the requirements can be quickly (and cheaply) identified. This is exactly what BDD proposes, the difference here is that the graphical representation is used to give the user the option of running the business process step by step and see how the system is behaving.

In other words, the user can check a live process and the live system that answers to this given process. If anything goes wrong, the tool will identify what and where, making corrections easier.

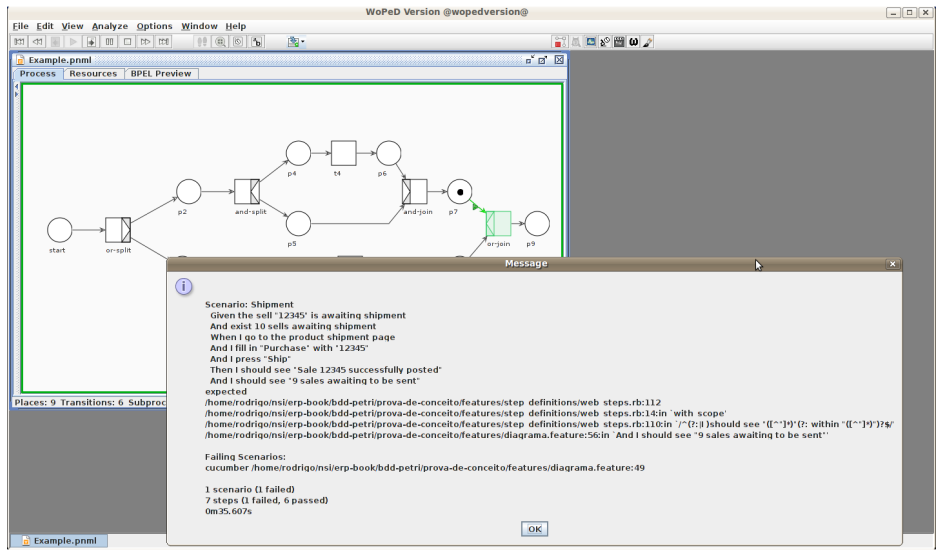

Figure 10: Error generated by code not compliant to requirements

## 4. Conclusions and Future Work

This paper aimed at describing in more detail the proposal originally presented by [4] and refined by [5]. Business Language Driven Development aims at simultaneously benefit from both visual representations of the processes and the security that they are all implemented in the right way through automated tests. It is important to note that BDD is a relatively new technique, subject to evolution, therefore, parts of the BLDD proposal are subject to re-interpretation.

Through the application of BLDD it is possible to reduce risks, costs and effort to develop, adapt, or customize Enterprise Information Systems such as ERP and CRM. Given an ERP framework wrapped by BLDD, it is possible to inject changes into the business process representations and, by running the associated tests, verify the exact points where the system needs to be changed to make it compliant to the changes. After implementing the changes (and their respective tests, of course), it is just a question of running the test set again to check if the new business requirements are correctly implemented.

A series of developments can follow the basic mappings here presented, such as describing them by Formal Methods and implementing the tools originally pointed by [4], as well as others:

-A tool for parsing a XML representation of the business process and generating the equivalent BDD steps, including the textual representation (GWT) of the process.

-An extension to the Eclipse's BPMN plugin so that modeling and programming can be done in the same environment. This tool would provide a type of Token Game for BPMN models.

-Investigation on how [models@run.time](models@run.time) [11] can help implementing the proof of concept here presented and also making the system even more adaptable.

-Investigation on the use of BLDD to help implementing Business Process Composition in a Service Oriented Architecture environment.

-A translator that, given the GWT constructs, generates the corresponding graphical representation.

-Define for the various notations the necessary steps, as shown in Code 2.

These tools should be able to handle the main BPM "languages" such as Petri Nets, BPMN, and UML Statechart and Activity Diagrams.

The next step of this proposal is to provide a complete analysis of the BDD process and map it to a new process that complements it by using graphical notations.